\documentclass[seceq]{ptptex}

\newcommand{\e}{\mathrm{e}}

\newcommand{\SL}{S_\Lambda}
\newcommand{\SFL}{S_{F,\Lambda}}
\newcommand{\SIL}{S_{I,\Lambda}}
\newcommand{\vev}[1]{\left\langle #1 \right\rangle}
\newcommand{\Tr}{\mathrm{Tr}~}
\newcommand{\K}[1]{K\left(#1/\Lambda\right)}

\newcommand{\Ld}[1]{\frac{\overrightarrow{\delta}}{\delta #1}}
\newcommand{\Rd}[1]{\frac{\overleftarrow{\delta}}{\delta #1}}
\newcommand{\lb}{\left\lbrace}
\newcommand{\rb}{\right\rbrace}
\newcommand{\Op}{\mathcal{O}}
\newcommand{\N}{\mathcal{N}}
\newcommand{\V}{\mathcal{V}}
\newcommand{\nn}{\nonumber}

\title{
An elementary proof of the non-renormalization theorem \\for the Wess-Zumino
model}

\author{
Hidenori \textsc{Sonoda}${}^1$
and Kayhan \textsc{\"Ulker}${}^2$
}
\inst{
${}^1$ Physics Department, Kobe University, Kobe 657-8501, Japan\\
${}^2$ Feza G\"ursey Research Institute, Istanbul, Turkey
}

\abst{
  Using the exact renormalization group (ERG) differential equation,
  we give an elementary proof of the non-renormalization theorem for
  the Wess-Zumino model.  We introduce auxiliary fields to linearize
  the supersymmetry transformation, but we do not rely on the
  superfield techniques.  We give sufficient background material on
  the Wilson action and the ERG formalism to make the paper
  self-contained.}

\begin{document}

\maketitle

\section{Introduction\label{introduction}}

The purpose of this paper is to prove the non-renormalization theorem
for the Wess-Zumino model \cite{Wess:1973kz} using an elementary method.
Here, what we call the non-renormalization theorem is the absence of
radiative corrections to the superpotential.  Two of its important
implications are that the expectation value of the scalar field is the
same as at the tree level, and that the beta function and anomalous
dimensions are related.  We will sketch a derivation of the latter in
sect.~\ref{implications}.

The cancellation of UV divergences in the Wess-Zumino model was first
analyzed in \citen{Iliopoulos:1974zv} shortly after the model was
introduced.  With the development of superfield \& supergraph
techniques in superspace it became possible to display all
simplifications due to supersymmetry
\cite{Fujikawa:1974ay,Grisaru:1979wc,Grisaru:1982zh,Wess:1992cp}.
A proof of the non-renormalization theorem was given in full details
for the first time in \citen{Grisaru:1979wc}, which relies on elaborate
uses of the superfield techniques. The simplest proof is the one given
by Seiberg \cite{Seiberg:1993vc, Seiberg:1994bp}, where the coupling
constants are promoted to chiral or antichiral constant superfields.
Our elementary proof does not rely on superfields, but relies on the
formulation of field theory using the exact renormalization group
(ERG) differential equation
\cite{Wilson:1973jj, Polchinski:1983gv}.  Though there is nothing
fancy about the ERG formalism, its familiarity cannot still be taken
for granted, and we give sufficient background in sects.~\ref{cutoff}
\& \ref{wilson}.

Our proof, to be given in sect.~\ref{proof}, is by no means the first
given within the ERG formalism.  The amalgamation of the superfield
and ERG formalisms was obtained in \citen{Pernici:1998ex} and
\citen{Bonini:1998ec}, for which the result of \citen{Grisaru:1979wc}
directly applies to proving the non-renormalization of the
superpotential.  A recent work by Rosten \cite{Rosten:2008ih} also
uses ERG to prove the non-renormalization theorem without using the
result of \citen{Grisaru:1979wc}.  Since our proof closely resembles
Rosten's (sect.~V of \citen{Rosten:2008ih}), we feel obliged to make
comparisons in order to justify the publication of the present paper.
In \citen{Rosten:2008ih}, a more general formulation of ERG is used to
study theories of a scalar chiral superfield, not necessarily
renormalizable perturbatively.  Especially, the use of Fourier
transforms in superspace makes his proof of non-renormalization less
transparent to follow compared with our work. Our proof is simpler,
depending only on the linearization of the supersymmetry
transformation by auxiliary fields, and it makes the advantage of the
ERG formalism more explicit.  It is a main goal of the present paper
to show that ERG provides such a straightforward formulation of
renormalizable theories.

We give a definite definition of the Wilson action $\SL$ with momentum
cutoff $\Lambda$.  The non-renormalization theorem we prove applies
only to $\SL$ with strictly positive $\Lambda$.  We use the four
dimensional euclidean space throughout the paper.

\section{The classical action and its symmetry\label{classical}}

The classical action of the Wess-Zumino model, without auxiliary
fields, is given by
\begin{eqnarray}
S'_{cl} &=& - \int d^4 x \bigg[ \partial_\mu \bar{\phi} \partial_\mu
\phi + |m|^2 |\phi|^2 
 + \bar{\chi}_L \sigma \cdot \partial \chi_R +  \frac{m}{2}
  \bar{\chi}_R \chi_R + \frac{\bar{m}}{2} \bar{\chi}_L \chi_L \nn\\
&& +  g \phi \frac{1}{2}
  \bar{\chi}_R \chi_R + \bar{g} \bar{\phi} \frac{1}{2} \bar{\chi}_L
  \chi_L 
 + \frac{|g|^2}{4} |\phi|^4 + m \phi \frac{\bar{g}}{2} \bar{\phi}^2
+ \bar{m} \bar{\phi} \frac{g}{2} \phi^2 \bigg],
\end{eqnarray}
where $m$ is a complex mass parameter, and $g$ is a complex
dimensionless coupling.  We use a bar to denote complex conjugation
except for the right- or left-handed two-component spinors
$\chi_{R,L}$, for which
\begin{equation}
\bar{\chi}_{R,L} \equiv \chi_{R,L}^T \sigma_y .
\end{equation}
We also define
\begin{equation}
\sigma_\mu \equiv (\vec{\sigma}, i),\quad
\bar{\sigma}_\mu \equiv (\vec{\sigma}, - i).
\qquad (\mu = 1, \cdots, 4)
\end{equation}

Shifting $\phi$ by $v \equiv \frac{m}{g}$, we obtain \cite{Tsao:1974nz}
\begin{eqnarray}
S'_{cl} &=& - \int d^4 x \Big[ \partial_\mu \bar{\phi} \partial_\mu \phi +
  \bar{\chi}_L \sigma \cdot \partial \chi_R \nn\\
&& + g \phi \frac{1}{2}
  \bar{\chi}_R \chi_R + \bar{g} \bar{\phi} \frac{1}{2} \bar{\chi}_L
  \chi_L 
 + \frac{|g|^2}{4} \left( \phi^2 - v^2 \right) \left(
    \bar{\phi}^2 - \bar{v}^2 \right) \Big].
\end{eqnarray}
With the shift, the action is invariant under the $\mathbf{Z}_2$
transformation:
\begin{equation}
\phi \to - \phi,\quad \bar{\phi} \to - \bar{\phi},\quad
\chi_R \to i \chi_R,\quad \chi_L \to - i \chi_L.
\end{equation}
This symmetry is broken spontaneously by the non-vanishing VEV
$\vev{\phi} = v$.  

Using complex auxiliary fields $F, \bar{F}$, we obtain an equivalent
action
\begin{eqnarray}
S_{cl} &\equiv& S'_{cl} 
 - \int d^4 x\, 
 \lb \bar{F} + i \frac{g}{2} \left( \phi^2 -
        v^2 \right) \rb \lb F + i \frac{\bar{g}}{2} \left(
        \bar{\phi}^2 - \bar{v}^2 \right)\rb \nn\\
&=& - \int d^4 x \bigg[ \partial_\mu \bar{\phi} \partial_\mu \phi +
  \bar{\chi}_L \sigma \cdot \partial \chi_R + \bar{F} F\nn\\
&&\quad + \frac{g}{2} \left( \phi 
  \bar{\chi}_R \chi_R + i F \phi^2 \right)
+ \frac{\bar{g}}{2} \left( \bar{\phi} \bar{\chi}_L \chi_L
  + i \bar{F} \bar{\phi}^2 \right)
 - i F \frac{g v^2}{2} - i \bar{F} \frac{\bar{g} \bar{v}^2}{2} \bigg].
\end{eqnarray}
The unfamiliar factors of the imaginary $i$ are due to our use of the
euclidean metric.  The action $S_{cl}$, written as above, depends on the
dimensionless $g$, the squared mass parameter $v^2$, and their
complex conjugates.  

The classical action is invariant under the following linear $N=1$
supersymmetry transformation:
\begin{equation}
\label{susy}
\begin{array}{c@{~=~}l@{\quad}c@{~=~}l}
\delta \phi & \bar{\chi}_R \xi_R\,,&
\delta \bar{\phi}& \bar{\chi}_L \xi_L,\\
\delta \chi_R & \partial_\mu \phi \bar{\sigma}_\mu \xi_L - i F \xi_R\,,&
\delta \chi_L & \partial_\mu \bar{\phi} \sigma_\mu \xi_R - i
\bar{F} \xi_L,\\
\delta F & - i \partial_\mu \bar{\chi}_R \bar{\sigma}_\mu \xi_L\,,&
\delta \bar{F} & - i \partial_\mu \bar{\chi}_L \sigma_\mu \xi_R,
\end{array}
\end{equation}
where $\xi_{R,L}$ are anticommuting constant spinors.  We note that
the chiral fields $(\phi, \chi_R, F)$ and the antichiral fields
$(\bar{\phi}, \chi_L, \bar{F})$ do not mix with each other under the
supersymmetry transformation.  

The parameters $g$ and $v$ are both complex, but their phases are
unphysical \cite{Seiberg:1994bp}. To see this, we consider the phase
changes
\begin{equation}
\begin{array}{c@{~\longrightarrow~}l@{\quad}
c@{~\longrightarrow~}l}
g & \e^{i \alpha} g,& \bar{g}& \e^{- i \alpha} \bar{g},\\
v & \e^{i \beta} v,& \bar{v}& \e^{- i \beta} \bar{v}.
\end{array}
\label{gvphasechange}
\end{equation}
The action $S_{cl}$ remains invariant, if we change the phases of the
fields at the same time as follows:
\begin{equation}
\begin{array}{c@{~\longrightarrow~}l@{\quad}
c@{~\longrightarrow~}l}
\phi& \e^{i \beta} \phi,& \bar{\phi}& \e^{- i \beta} \bar{\phi},\\
\chi_R& \e^{- i \frac{\alpha+\beta}{2}} \chi_R,&
\chi_L& \e^{i \frac{\alpha+\beta}{2}} \chi_L,\\
F & \e^{- i (\alpha + 2 \beta)} F,&
\bar{F}& \e^{ i (\alpha + 2 \beta)} \bar{F}.
\end{array}
\label{alphabeta}
\end{equation}
The transformation (\ref{alphabeta}) with $\alpha=0$ is the R-symmetry
transformation with R-character $\frac{1}{3}$, and that with
$\alpha+2\beta = 0$ is the R-symmetry transformation with R-character
$1$ \cite{Wess:1992cp}. The cubic interaction terms are invariant under
the former, and the terms linear in $F, \bar{F}$ are invariant under
the latter.  With $g$ and $v$ both non-vanishing, (\ref{alphabeta}) is
not a symmetry transformation.

\section{Momentum cutoff\label{cutoff}}

We regularize the theory using a smooth momentum cutoff.  We split the
bare action into the free and interaction parts:
\begin{equation}
S_B = S_{F,B} + S_{I,B}.
\label{bare action}
\end{equation}
We impose that each of $S_{F,B}$ and $S_{I,B}$ be invariant under the
supersymmetry transformation (\ref{susy}).  The free part is given by
\begin{equation}
S_{F,B} \equiv - \int_p \frac{1}{K\left(p/\Lambda_0\right)} \Big( p^2
    \bar{\phi} (-p) \phi (p)
  + \bar{\chi}_L (-p) i \sigma \cdot p
    \chi_R (p) + \bar{F} (-p) F(p) \Big),
\end{equation}
where $\int_p$ is a short-hand notation for $\int d^4 p/(2\pi)^4$, and
$K(p)$ is a smooth cutoff function satisfying the three properties:
\begin{enumerate}
\item $K(p)$ is a smooth non-negative decreasing function of $p^2$.
\item $K(p)$ vanishes as $p^2 \to \infty$ fast enough for UV
    finiteness.  (In our case, it suffices that $K$ vanishes as fast as
    $\frac{1}{p^4}$.) 
  \item $K(p) = 1$ for $p^2 < 1$.
\end{enumerate}
The free action implies the following propagators:
\begin{eqnarray}
\vev{\phi (p) \bar{\phi} (-p)}_{S_{F,B}} &=&
\frac{K(p/\Lambda_0)}{p^2},\\
\vev{\chi_R (p) \bar{\chi}_L (-p)}_{S_{F,B}} &=&
\frac{K(p/\Lambda_0)}{p^2} (-i) p \cdot \bar{\sigma},\\
\vev{F (p) \bar{F} (-p)}_{S_{F,B}} &=& K (p/\Lambda_0).
\end{eqnarray}
Thanks to the cutoff function $K$, the propagation of high momentum
modes are suppressed strongly, and the UV divergences are regularized.

The invariance under (\ref{susy}) and perturbative renormalizability
constrain the interaction action $S_{I,B}$ to have the following form:
\begin{eqnarray}
S_{I,B} &=& \int d^4 x \Big[ z_2\,
\lb \partial_\mu \bar{\phi} \partial_\mu \phi + \bar{\chi}_L \sigma
\cdot \partial \chi_R + \bar{F} F \rb\nn\\
&&\qquad + \left( -1 + z_3  \right) \lb
\frac{g}{2} \left( \phi \bar{\chi}_R \chi_R + i F \phi^2 \right) 
+ \frac{\bar{g}}{2} \left( \bar{\phi} \bar{\chi}_L \chi_L + i \bar{F}
\bar{\phi}^2 \right) \rb\nn\\
&&\qquad + \frac{g v^2}{2} i F + \frac{\bar{g} \bar{v}^2}{2} i \bar{F} \Big].
\end{eqnarray}
The renormalization constants $z_2$ \& $z_3$ depend on the cutoff
$\Lambda_0$ logarithmically.  The invariance under the R-symmetry
transformations (\ref{gvphasechange}) \& (\ref{alphabeta}) implies that
both $z_2$ and $z_3$ are functions of $|g|^2$.  The last two terms,
linear in $F$ or $\bar{F}$, give UV finite tadpoles, and they are not
renormalized \footnote{We may regard this as part of the
  non-renormalization theorem.}.

The non-renormalization theorem is that
\begin{equation}
z_3 = 0.
\label{nonrenormalization}
\end{equation}
We wish to prove this using the ERG differential equation, which we
will introduce in the next section.

\section{Wilson action\label{wilson}}

We renormalize the theory by examining the Wilson action with a
non-vanishing but finite cutoff $\Lambda$, instead of examining the
correlation functions.  Roughly speaking, the Wilson action is
obtained from the bare action with UV cutoff $\Lambda_0$ by
integration over the momentum modes above the IR cutoff $\Lambda$.
More precisely, the Wilson action is defined as follows:
\begin{equation}
\SL \equiv \SFL + \SIL,
\end{equation}
where the free part is the same as $S_{F,B}$ except $\Lambda_0$ is
replaced by $\Lambda$:
\begin{equation}
\SFL \equiv - \int_p \frac{1}{K\left(p/\Lambda\right)} \left( p^2
    \bar{\phi} (-p) \phi (p)
  + \bar{\chi}_L (-p) i \sigma \cdot p
    \chi_R (p) + \bar{F} (-p) F(p) \right).
\end{equation}
This is supersymmetric on its own.  The interaction part is defined by
\begin{eqnarray}
&&\exp \left[ \SIL [\phi, \cdots, \bar{F}] \right]
\equiv \int [d\phi' d\bar{\phi}'][d \chi'_R d\chi'_L ][dF' d\bar{F}']\,
\exp \Bigg[ \nn\\
&& \quad  - \int_p \frac{1}{K(p/\Lambda_0) -
  K(p/\Lambda)} \left( p^2
    \bar{\phi}' (-p) \phi' (p)
  + \bar{\chi}'_L (-p) i \sigma \cdot p
    \chi'_R (p) + \bar{F}' (-p) F'(p) \right) \nn\\
&&\qquad\qquad + S_{I,B} [\phi+\phi',\cdots,\bar{F}+\bar{F}'] \Bigg].
\label{SILintegralformula}
\end{eqnarray}
In terms of Feynman graphs, $\SIL$ consists of connected graphs with
the elementary three-point vertices and the propagators multiplied by
the cutoff function $K(p/\Lambda_0) - K(p/\Lambda)$, which is
approximately $1$ for $\Lambda^2 < p^2 < \Lambda_0^2$, and zero for
$p^2 < \Lambda^2$.  Note that $\SIL$ is not only supersymmetric, but
also invariant under the R-symmetry transformations
(\ref{gvphasechange}) \& (\ref{alphabeta}).

Alternatively, we can define $\SIL$ by the ERG differential equation
\begin{eqnarray}
&&- \Lambda \frac{\partial}{\partial \Lambda} \SIL = \int_q
\frac{\Delta (q/\Lambda)}{q^2} \left[
\frac{\delta \SIL}{\delta \phi (q)} \frac{\delta \SIL}{\delta \bar{\phi}
  (-q)} + 
\frac{\delta^2 \SIL}{\delta \phi (q) \delta \bar{\phi} (-q)} \right.\nn\\
&&\quad + \SIL \Rd{\chi_R (q)} (-i) q \cdot \bar{\sigma}
\Ld{\bar{\chi}_L (-q)} \SIL 
- \Tr  \Ld{\bar{\chi}_L (-q)} \SIL  \Rd{\chi_R (q)} (-i) q \cdot
\bar{\sigma}\nn \\
&&\left.\qquad\qquad + q^2 \frac{\delta \SIL}{\delta F(q)} \frac{\delta
      \SIL}{\delta \bar{F} 
  (-q)} + q^2 \frac{\delta^2 \SIL}{\delta F(q) \delta \bar{F} (-q)}
\right],
\label{ERGdiffeq}
\end{eqnarray}
where
\begin{equation}
\Delta (q) \equiv - 2 q^2 \frac{d}{dq^2} K(q),
\end{equation}
and by the initial condition
\begin{equation}
\SIL\Big|_{\Lambda = \Lambda_0} = S_{I,B}.
\label{initial}
\end{equation}
Note that $\Delta (q) = 0$ for $q^2 < 1$; hence, the $q$ in the ERG
differential equation is integrated over $q^2 > \Lambda^2$.  We can
regard (\ref{SILintegralformula}) as the integral formula of
(\ref{ERGdiffeq}) \& (\ref{initial}).

The dependence of $\SIL$ on the squared mass parameter $v^2$ and its
complex conjugate can be obtained trivially.  For this, we note that
if $\SIL$ is a solution of (\ref{ERGdiffeq}), so is
\begin{equation}
\tilde{S}_{I, \Lambda} \equiv \SIL - i \frac{g v^2}{2} F(0) - i \frac{\bar{g}
  \bar{v}^2}{2} \bar{F} (0),
\label{SILmassless}
\end{equation}
where 
\begin{equation}
F(0) \equiv \int d^4 x\, F(x),\quad
\bar{F} (0) \equiv \int d^4 x\, \bar{F} (x).
\end{equation}
Hence, $\tilde{S}_{I,\Lambda}$ is the Wilson action of the massless
theory, corresponding to $v^2=0$.

Let us now expand $\tilde{S}_{I, \Lambda}$ in powers of fields.  The
supersymmetry and the invariance under the R-symmetry transformations
(\ref{gvphasechange}) \& (\ref{alphabeta}) imply
\begin{eqnarray}
\tilde{S}_{I, \Lambda} &=& \int_p \V_2 (p) \lb p^2 \bar{\phi} (-p) \phi (p) +
\bar{\chi}_L (-p) i \sigma \cdot p \chi_R (p) + \bar{F} (-p) F(p)
\rb\nn\\
&& + \int_{p_1,p_2,p_3} \V_3 (p_1,p_2,p_3) \,(2\pi)^4 \delta^{(4)}
(p_1+p_2+p_3)\nn\\
&&\quad \times \lb \frac{g}{2} \left( \phi (p_1)
  \bar{\chi}_R (p_2) \chi_R (p_3) + i F (p_1) \phi (p_2) \phi (p_3) \right)
\right.\nn\\
&&\qquad + \left.
\frac{\bar{g}}{2} \left( \bar{\phi} (p_1) \bar{\chi}_L (p_2) \chi_L (p_3)
  + i \bar{F} (p_1) \bar{\phi}(p_2) \bar{\phi} (p_3) \right) \rb
\, + \,\cdots
\end{eqnarray}
up to terms cubic in the fields.  The coefficients $\V_2$ and $\V_3$
are scalar functions dependent on $|g|^2$.  $\V_3$ is symmetric with
respect to the three momenta.

Expanding $\V_2 (p)$ in powers of $\frac{p^2}{\Lambda^2}$, we obtain
\begin{equation}
\V_2 (p) = c_2 (\ln \Lambda/\mu) + \cdots .
\end{equation}
Similarly, expanding $\V_3 (p_1,p_2,p_3)$, we obtain
\begin{equation}
\V_3 (p_1,p_2,p_3) = - 1 + c_3 (\ln \Lambda/\mu) + \cdots .
\end{equation}
Both $c_2$ and $c_3$ are momentum independent constants, but they
depend on the cutoff $\Lambda$ logarithmically.  We have introduced an
arbitrary momentum scale $\mu$ to make the argument of the log
dimensionless.  The initial condition (\ref{initial}) implies
\begin{equation}
c_2 (\ln \Lambda_0/\mu) = z_2\,,\quad
c_3 (\ln \Lambda_0/\mu) = z_3\,.
\label{cinitial}
\end{equation}

Now, what about renormalization?  We tune the renormalization
constants $z_2, z_3$ so that the limit of the Wilson action
\begin{equation}
\lim_{\Lambda_0 \to \infty} \SIL
\end{equation}
exists order by order in $|g|^2$.  More concretely, we choose $z_2$,
$z_3$ such that
\begin{equation}
c_2 (0) = c_3 (0) = 0
\label{catzero}
\end{equation}
are satisfied.  This is a particular renormalization condition, where
$\mu$ plays the role of a renormalization scale.  To prove the
non-renormalization theorem (\ref{nonrenormalization}), it suffices to
show
\begin{equation}
- \Lambda \frac{\partial}{\partial \Lambda} c_3 (\ln \Lambda/\mu) = 0.
\label{c3nochange}
\end{equation}
Then, (\ref{cinitial}) and (\ref{catzero}) imply
\begin{equation}
c_3 (\ln \Lambda/\mu) = 0.
\label{c3-nonrenormalization}
\end{equation}
This is an alternative form of the non-renormalization theorem
(\ref{nonrenormalization}).

\section{Proof of the non-renormalization theorem\label{proof}}

To prove (\ref{c3nochange}), we examine the part of the Wilson action
proportional to
\begin{equation}
g \left( i F(0) \phi (0) \phi (0) + \phi (0) \bar{\chi}_R (0) \chi_R
  (0) \right),
\end{equation}
where all the fields are evaluated at zero momentum.  Its
coefficient is $-1+c_3 (\ln \Lambda/\mu)$.  The ERG differential
equation (\ref{ERGdiffeq}) gives
\begin{eqnarray}
&&- \Lambda \frac{\partial}{\partial \Lambda} c_3 (\ln \Lambda/\mu) \cdot
g \left( i F(0) \phi (0) \phi (0) + \phi (0) \bar{\chi}_R (0) \chi_R
  (0) \right)
=  \int_q \frac{\Delta (q/\Lambda)}{q^2} \label{dc3}\\
&& \times \lb
\frac{\delta^2}{\delta \phi (q)\delta \bar{\phi} (-q)} \V_5
- \Tr \Ld{\bar{\chi}_L (-q)} \V_5 \Rd{\chi_R (q)} (-i) q \cdot \bar{\sigma}
+ q^2 \frac{\delta^2}{\delta \bar{F}(-q) \delta F(q)} \V_5 \rb ,\nn
\end{eqnarray}
where $\V_5$ is the part of $\SIL$ proportional to one chiral field
with momentum $q$, one antichiral field with momentum $-q$, and three
chiral fields with zero momentum.  Since $\Delta (q/\Lambda) = 0$ for
$q^2 < \Lambda^2$, we can restrict $q^2 > \Lambda^2$ in $\V_5$.  We
enumerate all possibilities for $\V_5$ that satisfy the constraints of
supersymmetry and the invariance under the R-symmetry transformations
(\ref{gvphasechange}) \& (\ref{alphabeta}).  The most general $\V_5$
is given in the following form:
\begin{eqnarray}
\V_5 &=& g \int_q \left[
\lb \bar{\chi}_R (-q) (-i) q \cdot \bar{\sigma} \chi_L (q) + i F (-q)
i \bar{F} (q) - q^2 \bar{\phi}(q) \phi (-q) \rb\right.\nn\\
&&\qquad\qquad \times \frac{1}{2} \lb w_1 (q) \phi (0) \phi (0) i F(0) + w_2
    (q) \bar{\chi}_R (0) 
    \chi_R (0) \phi (0) \rb\nn\\
&&\quad + \left(w_1 (q) - w_2(q)\right) i F(0) \lb
\phi (0) \left( \bar{\chi}_R (0) (-i) q \cdot \bar{\sigma} \chi_L (q)
    + i F(0) i \bar{F} (q) \right) \phi (-q) \right.\nn\\
&&\qquad\qquad \left.\left.+ \left(\phi (-q) \frac{1}{2} \bar{\chi}_R
            (0) \chi_R (0) + \phi (0)
        \bar{\chi}_R (-q) \chi_R (0) \right) i \bar{F} (q) \rb \right]\,,
\label{quintic}
\end{eqnarray}
where $w_1 (q), w_2 (q)$ are arbitrary functions of $q^2$ and $|g|^2$.
For example, at 1-loop, we obtain
\begin{equation}
\lb\begin{array}{c@{~=~}l}
w_1 (q) & 2 |g|^4 \frac{1-\K{q}}{q^2} \int_r \frac{(1-\K{(r+q)})^2
  (1-\K{r})^2}{r^4 (r+q)^4} \, r^2,\\
w_2 (q) & |g|^4 \frac{1 - \K{q}}{q^2} \int_r \frac{(1-\K{(r+q)})^2
  (1-\K{r})^2}{r^4 (r+q)^4} \, \left( r^2 - \frac{q^2}{2}\right).
\end{array}\right.
\end{equation}
Substituting (\ref{quintic}) into (\ref{dc3}), we can verify
(\ref{c3nochange}) by simple algebra.  Please see Appendix
\ref{susy-invariants} for an explanation of this vanishing and
a more systematic derivation of (\ref{quintic}).

\section{Generalization}

The non-renormalization theorem (\ref{nonrenormalization}) or
(\ref{c3-nonrenormalization}) can be generalized further using the
invariance of the Wilson action under the R-symmetry transformations
(\ref{gvphasechange}) \& (\ref{alphabeta}).

Let us first define the superpotential $V_\Lambda [\Phi] +
\bar{V}_\Lambda [\bar{\Phi}]$ using the Wilson action.  The
superpotential $V_\Lambda [\Phi]$ is the part of $\SIL$ that depends
only on the chiral fields $\Phi \equiv \lbrace \phi, \chi_R, F
\rbrace$ with no space derivatives.  Similarly, $\bar{V}_\Lambda
[\bar{\Phi}]$ is the part dependent only on $\bar{\Phi} \equiv \lbrace
\bar{\phi}, \chi_L, \bar{F}\rbrace$ with no space derivatives.
$\bar{V}_\Lambda$ is basically the complex conjugate of $V$, where we
interpret $\chi_L$ as the complex conjugate of $\chi_R$.  The
non-renormalization theorem (\ref{c3-nonrenormalization}) implies that
the cubic part of $V_\Lambda [\Phi]$ is given exactly by
\begin{equation}
- \frac{g}{2} \int d^4 x \, \lb i F \phi^2 + \phi \bar{\chi}_R \chi_R
\rb .
\end{equation}
Now, what about higher order terms?

We know that the higher order terms depend only on the chiral fields
$(\phi, \chi_R, F)$ and the coupling constants $g, \bar{g}$.  The most
general supersymmetric term consisting of $n (\ge 4)$ chiral fields is
given by
\[
g |g|^{2 l}\int d^4 x\,  \left( i F \phi^{n-1} + \frac{n-1}{2}
  \phi^{n-2} \bar{\chi}_R  \chi_R  \right) ,
\]
where $l$ is an arbitrary non-negative integer.  The first factor of
$g$ is for the invariance under the R-symmetry transformations
(\ref{gvphasechange}) \& (\ref{alphabeta}) with $\beta=0$.  But the
above is not invariant under the R-symmetry transformations
(\ref{gvphasechange}) and (\ref{alphabeta}) with $\beta \ne 0$.

Thus, we conclude that the chiral part of the superpotential is
exactly given by
\begin{equation}
  V_\Lambda [\Phi] = - \frac{g}{2} \int d^4 x \, \lb i F (\phi^2 - v^2) + \phi
  \bar{\chi}_R \chi_R \rb .
\label{superpotential}
\end{equation}

\section{Implications\label{implications}}

For completeness of the paper, we would like to give an argument that
relates the anomalous dimensions of the fields and $v^2$ to the beta
function of $|g|^2$.  The anomalous dimensions beyond 1-loop and the
beta function beyond 2-loop are scheme dependent, and the relations we
derive are valid only for the particular scheme adopted in the present
paper.  The anomalous dimensions and beta function can be derived from
the dependence of the Wilson action $\SL$ on the renormalization point
$\mu$ as discussed in \citen{Pernici:1998tp} and \citen{Sonoda:2006ai}.
In the following we give a hand-waving but more straightforward
derivation, relegating a systematic derivation to Appendix \ref{beta}.

We replace the bare action (\ref{bare action}) na\"ively by
\begin{eqnarray}
S_B &=& \int d^4 x \Bigg[ - Z \lb \partial_\mu \bar{\phi} \partial_\mu
  \phi + \bar{\chi}_L \sigma \cdot \partial \chi_R + \bar{F} F \rb\nn\\
&&\quad - Z^{\frac{3}{2}} Z_g \left( g \lb i F \phi^2 + \phi
  \bar{\chi}_R \chi_R \rb + \bar{g} \lb i \bar{F} \bar{\phi}^2 +
  \bar{\phi} \bar{\chi}_L \chi_L \rb \right)\nn\\
&&\quad + Z^{\frac{1}{2}} Z_g Z_{v^2} \lb \frac{g v^2}{2} i F +
\frac{\bar{g} \bar{v}^2}{2} i \bar{F} \rb \Bigg] ,
\end{eqnarray}
where
\begin{equation}
Z = 1 - c_2,\quad
Z^{\frac{3}{2}} Z_g = 1 - c_3,\quad
Z^{\frac{1}{2}} Z_g Z_{v^2} = 1.
\end{equation}
$Z$ is the renormalization constant for the fields, $g_B = Z_g g$ is the
bare coupling, and $v_B^2 = Z_{v^2} v^2$ is the bare squared mass parameter.
The non-renormalization theorem (\ref{nonrenormalization}) gives
\begin{equation}
Z = Z_g^{- \frac{2}{3}},\quad Z_{v^2} = Z.
\end{equation}
Defining the beta function $\beta$, the anomalous dimension
$\beta_{v^2}$ of $v^2$, and that of fields $\gamma$ by
\begin{equation}
\mu \frac{\partial}{\partial \mu} \ln Z_g = \frac{\beta}{2
  |g|^2} ,\quad
\mu \frac{\partial}{\partial \mu} \ln Z_{v^2} = \beta_{v^2},\quad
\mu \frac{\partial}{\partial \mu} \ln Z = 2 \gamma ,
\end{equation}
we obtain
\begin{equation}
\frac{\beta}{2 |g|^2} = - 3 \gamma,\quad
\beta_{v^2} = 2 \gamma ,
\end{equation}
as first found in \citen{Iliopoulos:1974zv}.

\section{Concluding remarks}

In this paper we have proven the non-renormalization theorem
(\ref{c3-nonrenormalization}) \& (\ref{superpotential}) by an
elementary use of ERG.  Even though our proof does not rely on the
superfield techniques, it still depends on the linearization of the
supersymmetry transformation via auxiliary fields.  The Wess-Zumino
model can be constructed without auxiliary fields \footnote{For a
  construction within the ERG formalism, see \citen{Sonoda:2008dz}.},
and it should be interesting to formulate and prove the
non-renormalization theorem without auxiliary fields.

\section*{Acknowledgements}
We thank Prof.~M.~Sakamoto for discussions.

\appendix

\section{Non-renormalization theorem in terms of component
  fields\label{susy-invariants}}

In the component field formalism, the algebraic source of the
non-renormalization theorems of supersymmetric theories is the
cohomological structure of the supersymmetry transformation: any
supersymmetric invariants can be written as multiple supervariations
of integrals over field polynomials \cite{Flume:1999jf,Ulker:2000ir}.

Let us write the N=1 supersymmetry transformation $\delta$ in terms of
left and right supersymmetry generators:
\begin{equation}
\delta = \bar{\xi}_L Q_L + \bar{\xi}_R Q_R. 
\end{equation}
Any supersymmetric invariant in four dimensions is the highest
component of the respective super multiplet, and it can be written as
\begin{equation}
I =  \bar Q_R  Q_R \bar  Q_L Q_L X + \bar Q_R Q_R Y + \bar Q_L  Q_L
  \bar Y, \label{decomposition}
\end{equation}
where $X$, $Y$, and $\bar{Y}$ are integrals over polynomials of fields
that belong to the respective supersymmetry multiplets.  $Y$ satisfies
$Q_L Y = 0$, and $\bar{Y}$ satisfies $Q_R \bar Y = 0$. But $X$ need
not to be invariant under $Q_L$ or $Q_R$.

For instance, let us construct the chiral part, $\bar Q_R Q_R Y$, for
the Wilson action of the Wess-Zumino model. By analyzing the
supersymmetry transformations of the fields, one finds that $Y$ can
only be written as (an integral of) $\phi^n$.  All the other
polynomials annihilated by $Q_L$ can be written as $\bar Q_L Q_L
\tilde X'$ so that these make part of the first term in
(\ref{decomposition}).  For instance, $\partial^2 \phi \cdot
\phi^{n-1}$ can be written as $\bar Q_L Q_L (F \phi^{n-1})$, and
$\phi^n \bar F^m$ as $\bar Q_L Q_L (\phi^n \bar \phi \bar F^{m-1})$.

It is now straightforward to rewrite Seiberg's proof of the
non-renormalization theorem \cite{Seiberg:1993vc, Seiberg:1994bp}
using component fields.  Following Seiberg \cite{Seiberg:1993vc,
  Seiberg:1994bp}, we promote the parameters $g, \bar{g}$ to constant
fields that belong to chiral multiplets with the R-characters
determined by (\ref{gvphasechange}). Then, a similar argument as above
shows that $Y$ is a polynomial of only $g$ and $\phi$.  Since the
action is R-symmetric, $Y$ must transform as $\e^{i(\alpha+3\beta)}$
under (\ref{gvphasechange}) \& (\ref{alphabeta}). This leaves only a
constant multiple of $\int d^4 x\,g \phi^3$.  Hence, the chiral
(antichiral) part of the action related with $Y$ ($\bar Y$) does not
receive radiative corrections.  This is exactly the same argument as
Seiberg's \cite{Seiberg:1993vc, Seiberg:1994bp}. Therefore, we
conclude that any radiative correction to the supersymmetric Wilson
action can be written as $\bar Q_R Q_R \bar Q_L Q_L X$.

Now, using the above construction of supersymmetric invariants, we can
easily derive $\V_5$ given by (\ref{quintic}). Note that the quintic
vertex $\V_5$ consists of terms of the form
 \begin{equation}
\Phi (q) \bar{\Phi} (-q) \cdot \Phi (0) \Phi (0) \Phi (0),
\end{equation}
where $\Phi$ and $\bar{\Phi}$ denote generic chiral and antichiral
component fields, and the mass dimension of $\Phi (0) \Phi (0) \Phi
(0)$ must be at most $5$.  Considering the R-symmetry
(\ref{gvphasechange}) \& (\ref{alphabeta}) and the dimensionality of
the fields, we conclude the following:
\begin{enumerate}
\item There is no possible $Y$ satisfying $Q_L Y =0$ that cannot be
  written as $\bar{Q}_L Q_L X$.
\item There are nine possible field polynomials for $X$, but under the
  action of
\[
\bar{Q}_L Q_L \bar{Q}_R Q_R,
\]
only the following two are independent:
\[
X= \lb g \bar \phi\phi \cdot \phi\bar \chi_R\chi_R \, ,\, g
\bar\phi \phi \cdot iF\phi^2 \rb .
\]
\end{enumerate}
It is then easy to show that (\ref{quintic}) can be obtained as
\begin{eqnarray}
  \mathcal{V}_5 &=& \frac{g}{8} \bar Q_R Q_R\bar Q_L Q_L \int_q \\
&& \left[ w_1 (q) \bar
    \phi(q)\phi(-q)  iF(0)\phi(0)\phi(0) + w_2 (q) \bar \phi(q)\phi(-q)
    \phi(0)\bar\chi_R(0)\chi_R(0) \right].\nn
\end{eqnarray}

Note that in Sec.V, $\V_5$ is found by enumerating all the
possibilities allowed by R-symmetry and dimensionality. However, the
above construction is more systematic and simpler.  Moreover,
(\ref{c3nochange}) comes as no surprise since the loop contraction
\begin{equation}
\int_q \frac{\Delta (q/\Lambda)}{q^2} \left( \frac{\delta^2}{\delta
    \phi (q) \delta \phi (-q)} + \cdots \right)
\end{equation}
is supersymmetric and commutes with $\bar Q_R Q_R\bar Q_L Q_L$.
Hence, the loop contraction gives $\bar Q_R Q_R\bar Q_L Q_L
iF(0)\phi(0)\phi(0)$ and $\bar Q_R Q_R\bar Q_L Q_L
\phi(0)\bar\chi_R(0)\chi_R(0)$, both of which vanish identically.

We wish to give more details elsewhere on the use of the component
field formalism for the cohomological construction and for the proof
of the non-renormalization theorem.

\section{The relation among $\beta$, $\gamma$, and
  $\beta_{v^2}$\label{beta}}

The derivation of the beta function and anomalous dimensions from the
Wilson action has been discussed in \citen{Pernici:1998tp} and
\citen{Sonoda:2006ai}.  Here, we follow the latter.

The $\mu$ dependence of the Wilson action can be written as
\begin{equation}
- \mu \frac{\partial}{\partial \mu} \SL = \frac{\beta}{2 |g|^2} \left( g
    \Op_g + \bar{g} \Op_{\bar{g}} \right) + \beta_{v^2} \left( v^2
    \Op_{v^2} + \bar{v}^2 \Op_{\bar{v}^2} \right) + \gamma \N ,
\label{mudep}
\end{equation}
where $\beta$, $\beta_{v^2}$, and $\gamma$ are functions of $|g|^2$,
and the composite operators $\Op_g$, $\Op_{v^2}$, $\N$ are defined by
\begin{eqnarray}
\Op_g &\equiv& - \partial_g \SIL,\\
\Op_{v^2} &\equiv& - \partial_{v^2} \SIL = - \frac{i}{2} g F (0),\\
\N &\equiv& - \int_p \K{p} \left[
[\phi] (p) \frac{\delta \SL}{\delta \phi (p)} + \frac{\delta}{\delta
  \phi (p)} [\phi] (p) + [\bar{\phi}] (p) \frac{\delta \SL}{\delta
  \bar{\phi} (p)} + \frac{\delta}{\delta \bar{\phi} (p)} [\bar{\phi}]
(p)\right.\nn\\
&& + \SL \Rd{\chi_R (p)} [\chi_R] (p) - \Tr [\chi_R ] (p) \Rd{\chi_R
  (p)} + \SL \Rd{\chi_L (p)} [\chi_L] (p) - \Tr [\chi_L] (p) \Rd{\chi_L
  (p)} \nn\\
&&\left.\, + [F] (p) \frac{\delta \SL}{\delta F(p)} + \frac{\delta}{\delta
  F(p)} [F] (p) +  [\bar{F}] (p) \frac{\delta \SL}{\delta \bar{F}(p)} +
\frac{\delta}{\delta \bar{F}(p)} [\bar{F}] (p) \right] ,
\end{eqnarray}
where the elementary fields in square brackets are defined by
\begin{equation}
[\phi] (p) \equiv \phi (p) + \frac{1 - K(p/\Lambda)}{p^2}
\frac{\delta \SIL}{\delta \bar{\phi} (-p)}, \,\cdots .
\end{equation}
The operator $\N$ is the equation of motion operator that counts the
number of fields:
\begin{equation}
\vev{\N \underbrace{\phi (p_1) \cdots }_{N\,\mathrm{fields}}} =
N \vev{\phi (p_1) \cdots }.
\end{equation}
$\beta$ is the beta function of $|g|^2$, $\beta_{v^2}$ is
the anomalous dimension of $v^2$, and $\gamma$ is the common anomalous
dimension of the matter fields, since
(\ref{mudep}) implies
\begin{equation}
  \left( - \mu \frac{\partial}{\partial \mu} + \frac{\beta}{2|g|^2}
    ( g \partial_g + \bar{g} \partial_{\bar{g}}) + \beta_{v^2} (v^2 \partial_{v^2}
    + \bar{v}^2 \partial_{\bar{v}^2} ) \right) \vev{ \phi (p_1) \cdots }
  = N \gamma \vev{ \phi (p_1) \cdots}.
\end{equation}

To extract the coefficients $\beta, \beta_{v^2}, \gamma$, we expand
the Wilson action in powers of fields, and expand the coefficients in
powers of momenta.  Alternatively, we can examine the asymptotic
behavior of the Wilson action for large
$\Lambda$\cite{Sonoda:2006ai}:
\begin{enumerate}
\item The asymptotic behavior of $-\mu \partial_\mu \SL$ is given by
\begin{equation}
- \mu \frac{\partial}{\partial \mu} \SL \stackrel{\Lambda \to
  \infty}{\longrightarrow} 
\Lambda \frac{\partial}{\partial \Lambda}
c_2 (\ln \Lambda/\mu) \int d^4 x \left( \partial_\mu
    \bar{\phi} \partial_\mu \phi + \bar{\chi}_L \sigma \cdot \partial
    \chi_R + \bar{F} F \right).
\end{equation}
\item The asymptotic behavior of the $g, \bar{g}$ derivatives is given by
\begin{eqnarray}
g \Op_g + \bar{g} \Op_{\bar{g}} &=& - g \partial_g \SL -
\bar{g} \partial_{\bar{g}} \SL\nn\\
&\stackrel{\Lambda \to
  \infty}{\longrightarrow}& \int d^4 x \, \left[ - 2 |g|^2
    \frac{\partial}{\partial |g|^2} c_2
    (\ln \Lambda/\mu) \left( \partial_\mu
    \bar{\phi} \partial_\mu \phi + \bar{\chi}_L \sigma \cdot \partial
    \chi_R + \bar{F} F \right)\right.\nn\\
&&\quad + \frac{1}{2} \left( g \phi \bar{\chi}_R \chi_R + g \phi^2 i F
+ \bar{g} \bar{\phi} \bar{\chi}_L \chi_L + \bar{g} \bar{\phi}^2 i
\bar{F} \right)\nn\\
&&\quad \left.- \frac{1}{2} \left( g v^2 i F + \bar{g} \bar{v}^2 i \bar{F}
\right) \right].
\end{eqnarray}
\item $\Op_{v^2}$ and $\Op_{\bar{v}^2}$ are exactly given by
\begin{equation}
v^2 \Op_{v^2} + \bar{v}^2 \Op_{\bar{v}^2} =
 - \frac{i}{2} \left( g v^2 F (0) + \bar{g}
    \bar{v}^2 \bar{F} (0) \right).
\end{equation}
\item The asymptotic behavior of the field counting operator is the
  most complicated:
\begin{eqnarray}
\N &\stackrel{\Lambda \to \infty}{\longrightarrow}& \int d^4 x \,
\left[
2 (1 - c_2) \left( \partial_\mu \bar{\phi} \partial_\mu \phi +
    \bar{\chi}_L \sigma \cdot \partial \chi_R + \bar{F} F \right)
\right.\nn\\
&&\quad + \frac{3}{2} \left( g \phi \bar{\chi}_R \chi_R + g i F \phi^2
    + \bar{g} \bar{\phi} \bar{\chi}_L \chi_L + \bar{g} i \bar{F}
    \bar{\phi}^2 \right)\nn\\
&&\quad \left.- \frac{1}{2} \left( g v^2 i F + \bar{g} \bar{v}^2 i \bar{F}
\right) \right]\\
&& - 2 \int_p \frac{\K{p} (1-\K{p})}{p^2} \left(
\frac{\delta^2 \SIL}{\delta \phi (p) \delta \bar{\phi} (-p)} \right.\nn\\
&&\quad \left.+ \Tr
(-i) p \cdot \bar{\sigma} \Ld{\bar{\chi}_L (-p)} \SIL \Rd{\chi_R (p)} 
+ p^2 \frac{\delta^2 \SIL}{\delta \bar{F} (-p) \delta F (p)} \right) ,\nn
\end{eqnarray}
where the loop momentum $p$ is of order $\Lambda$.  For the loop
integral, we need the part of $\SIL$ including a chiral field of
momentum $p$, an antichiral field of momentum $-p$, and a number of
fields with momenta low compared with $\Lambda$.  The part including
three chiral fields at zero momentum vanishes as explained in the
proof of the non-renormalization theorem.  Only the part proportional
to the kinetic term survives the loop integral.
\end{enumerate}

Hence, we obtain
\begin{eqnarray}
&& \frac{\beta}{2 |g|^2} \left( g
    \Op_g + \bar{g} \Op_{\bar{g}} \right) + \beta_{v^2} \left( v^2
    \Op_{v^2} + \bar{v}^2 \Op_{\bar{v}^2} \right) + \gamma \N\nn\\
&\stackrel{\Lambda \to \infty}{\longrightarrow}&
\int d^4 x \Bigg[ \left( - \beta \frac{\partial c_2}{\partial |g|^2} 
+ 2 \gamma (1 - c_2) - 2 \gamma \int_p \frac{\K{p} (1 - \K{p})}{p^2}
B_4 (0,p)\right) \nn\\
&&\qquad\qquad \cdot \left(\partial_\mu \bar{\phi} \partial_\mu \phi +
    \bar{\chi}_L \sigma \cdot \partial \chi_R + \bar{F} F \right)\nn\\
&&\qquad + \left( \frac{\beta}{2|g|^2} + 3 \gamma \right)
 \cdot \frac{1}{2} \left( g \phi \bar{\chi}_R \chi_R + g i F \phi^2
    + \bar{g} \bar{\phi} \bar{\chi}_L \chi_L + \bar{g} i \bar{F}
    \bar{\phi}^2 \right) \nn\\
&&\qquad + \left( - \frac{\beta}{2|g|^2}  - \beta_{v^2} -
    \gamma\right) \frac{1}{2} \left( g v^2 i F + \bar{g} \bar{v}^2 i
    \bar{F} \right) \Bigg],
\end{eqnarray}
where $B(0,p)$ is the coefficient of a term in $\SIL$, containing a
pair of chiral and antichiral fields with momentum of order $\Lambda$
and a pair with low momentum.  Thus, we obtain the desired relation
\begin{equation}
\frac{\beta}{2 |g|^2} = - 3 \gamma,\quad
\beta_{v^2} = 2 \gamma,
\end{equation}
and an equation that determines the anomalous dimension $\gamma$:
\begin{equation}
\frac{\partial c_2 (\ln \Lambda/\mu)}{\partial \ln \Lambda/\mu}
=  6 \gamma |g|^2 \frac{\partial c_2}{\partial |g|^2} 
+ 2 \gamma (1 - c_2) - 2 \gamma \int_p \frac{\K{p} (1 - \K{p})}{p^2}
B_4 (0,p).
\end{equation}
We can calculate $\gamma$ by computing $c_2$ and $B_4$ perturbatively.

Up to 2-loop, we obtain
\begin{equation}
\lb\begin{array}{c@{~=~}l}
B_4^{(0)} (0,p) & - |g|^2 \frac{1-K(p/\Lambda)}{p^2},\\
c_2^{(1)} (\ln \Lambda/\mu) & \frac{|g|^2}{(4\pi)^2} \ln \frac{\Lambda}{\mu},\\
c_2^{(2)} (\ln \Lambda/\mu) & \frac{|g|^4}{(4 \pi)^4} \left[
\left(\ln \frac{\Lambda}{\mu}\right)^2 - a \ln \frac{\Lambda}{\mu} \right],
\end{array}\right.
\end{equation}
where the superscript denotes the number of loops, and the
constant $a$ is given by
\begin{eqnarray}
a &\equiv& (4 \pi)^4 \Bigg[ \frac{3}{2} \int_q \frac{\Delta (q)
      (1-K(q))^2}{q^4} \int_r \frac{1 - K(r)}{r^2} \left( \frac{1 -
          K(r+q)}{(r+q)^2} - \frac{1-K(r)}{r^2} \right) \nn\\
&&\quad + \int_q
    \frac{\Delta (q)}{q^2} \int_r \frac{(1-K(r))^3 (1-K(r+q))}{r^4
      (r+q)^2} \Bigg].
\end{eqnarray}
Note that the value of $a$ depends on the choice of the cutoff
function $K$.

Using the above, we obtain the following results for the anomalous dimension:
\begin{eqnarray}
2 \gamma^{(1)} &=& \dot{c}_2^{(1)} (\ln \Lambda/\mu) = |g|^2 \frac{1}{(4\pi)^2},\\
2 \gamma^{(2)} &=& \dot{c}_2^{(2)} (\ln \Lambda/\mu) - 4 \gamma^{(1)}
\cdot c_2^{(1)} - 2 \gamma^{(1)} |g|^2 \int_q \frac{K(1-K)^2}{q^4}\nn\\
&=& - \frac{|g|^4}{(4\pi)^4} b,
\end{eqnarray}
where
\begin{eqnarray}
b &\equiv& a + (4 \pi)^2 \int_q \frac{K(1-K)^2}{q^4}\nn\\
&=& (4\pi)^4 \Bigg[ \frac{3}{2} \int_q \frac{\Delta (q)
      (1-K(q))^2}{q^4} \int_r \frac{1 - K(r)}{r^2} \left( \frac{1 -
          K(r+q)}{(r+q)^2} - \frac{1-K(r)}{r^2} \right)\nn \\
&&\quad + \int_q
    \frac{\Delta (q)}{q^2} \int_r \frac{(1-K(r))^3 (1-K(r+q))}{r^4
      (r+q)^2} + \frac{1}{(4\pi)^2} \int_q \frac{K(1-K)^2}{q^4}\Bigg]\nn\\
&=& 1.
\end{eqnarray}
The value of $b$ is independent of the choice of $K$ \footnote{We have
  computed $b$ using the sharp cutoff function $K(q) = \theta
  (1-q^2)$.}.

Thus, up to 2-loop, we obtain
\begin{equation}
2 \gamma = \frac{|g|^2}{(4 \pi)^2} - \frac{|g|^4}{(4 \pi)^4}.
\end{equation}
This agrees with the known result \cite{Abbott:1980jk, Pernici:1998ex}.

\end{document}